\documentclass[aps,twocolumn,showpacs,floats,superscriptaddress]{revtex4}
\usepackage{graphicx}
\usepackage{amssymb,amsmath}
\begin{document}

\title{Phase diagram and thermodynamics of the three-dimensional Bose-Hubbard model}

\author{B. Capogrosso-Sansone}
\affiliation{Department of Physics, University of Massachusetts,
Amherst, MA 01003, USA}

\author{N.V. Prokof'ev}
\affiliation{Department of Physics, University of Massachusetts,
Amherst, MA 01003, USA} \affiliation{BEC-INFM, Dipartimento di
Fisica, Universit$\acute{a}$ di Trento, Via Sommarive 14, I-38050
Povo, Italy} \affiliation{Russian Research Center ``Kurchatov
Institute'', 123182 Moscow, Russia}

\author{B.V. Svistunov}
\affiliation{Department of Physics, University of Massachusetts,
Amherst, MA 01003, USA}
\affiliation{Russian Research Center ``Kurchatov Institute'',
123182 Moscow, Russia}


\begin{abstract}
We report results of quantum Monte Carlo simulations of the
Bose-Hubbard model in three dimensions. Critical parameters for
the superfluid-to-Mott-insulator transition are determined with
significantly higher accuracy than it has been done in the past.
In particular, the position of the critical point at filling
factor $n=1$ is found to be at $(U/t)_{\rm c} = 29.34(2)$, and the
insulating gap $\Delta$ is measured with accuracy of a few percent
of the hopping amplitude $t$. We obtain the effective mass of
particle and hole excitations in the insulating state---with
explicit demonstration of the emerging particle-hole symmetry and
relativistic dispersion law at the transition tip---along with the
sound velocity in the strongly correlated superfluid phase. These
parameters are the necessary ingredients to perform analytic
estimates of the low temperature ($T\ll \Delta$) thermodynamics in
macroscopic samples. We present accurate thermodynamic curves,
including these for specific heat and entropy, for typical
insulating ($U/t=40$) and superfluid ($t/U=0.0385$) phases. Our
data can serve as a basis for accurate experimental thermometry,
and a guide for appropriate initial conditions if one attempts to
use interacting bosons in quantum information processing.
\end{abstract}

\pacs{03.75.Hh, 03.75.Lm, 75.40.Mg}
\maketitle

\section{Introduction}

\parindent  8.mm

In the past decade, strongly correlated lattice quantum systems
have been attracting a lot of interest and effort. Remarkably,
simple yet nontrivial models which contain most of the important
many-body physics and known in the theory community for many years
can be now realized and studied experimentally. For the first time
theoretical predictions and experimental data for strongly
correlated states can be directly tested against each other in the
ideal setup when all model ingredients are known and controlled.

Experimentally, lattice systems are realized by trapping atoms in
an optical lattice, a periodic array of potential wells resulting
from the dipole coupling of the atoms to the electric field of the
standing electromagnetic wave produced by a laser. Optical
lattices are a very powerful and versatile tool. By changing the
laser parameters and configuration, the properties and geometry of
the optical lattice can be finely tuned \cite{Jaksch2}.
Ultimately, this results in the possibility of controlling the
Hamiltonian parameters and exploring various regimes of interest.
In particular, ultra-cold Bose atoms trapped in an optical lattice
are an experimental realization of the Bose-Hubbard model. The
model has been studied in the seminal paper by Fisher, Weichman,
Grinstein, and Fisher, Ref.~\onlinecite{Fisher}, and its physical
realization with ultra-cold atoms trapped in an optical lattice
has been envisioned in Ref.~\cite{Jaksch}. Few years later, the
Bose-Hubbard system was produced in the laboratory
~\cite{Greiner}. Since then, the field remains very active
\cite{Batrouni, Bloch_theor, Isacsson, Bloch_spatial, Gerbier1,
Clark}, not only because theoretical predictions and experimental
techniques still have to be substantially improved to claim the
quantitative agreement, but also because of the new physical
applications.

At zero temperature, a system of bosons with commensurate filling
factor undergoes a superfluid-to-Mott insulator (SF-MI) quantum
phase transition. The ground state of MI can be used in quantum
information processing to initialize a large set of qubits (the
main remaining challenge is in addressing single atoms to build
quantum gates, see Ref.~\cite{Jaksch2} and references therein).
Atomic systems in optical lattices have the advantage of being
well isolated from the environment. This results in a relatively
long decoherence time of the order of seconds ~\cite{Jaksch2} and
therefore the possibility of building long-lived entangled many body
states. These properties make MI groundstates good candidates for
building blocks of a quantum computer. Another possible
application is in interferometric measurements
\cite{interferometry}. It has been argued \cite{Burnett1,
Burnett2, Rodriguez} that using the superfluid-to-Mott-insulator
phase transition to entangle and disentangle atomic Bose-Einstein
condensate one can go beyond the Heisenberg-limited
interferometry.

A system of bosons with short-range repulsive pair interaction
trapped in an optical lattice is described by the Bose-Hubbard
Hamiltonian:
\begin{equation}
H\; =\; -t\sum_{<ij>} b^{\dag}_i\,b_j +\frac{U}{2}\sum_i
n_i(n_i-1) -\sum_i \mu_i n_i\; , \label{BH}
\end{equation}
where $b^{\dag}_i$  and $b_i$ are the bosonic creation and
annihilation operators on the site $i$, $ t$ is the hopping matrix
element, $U$ is the on-site repulsion and $\mu_i=\mu-V(i)$ is the
sum of the chemical potential $\mu$ and the confining potential
$V(i)$. In what follows, we consider bosons in the simple cubic
lattice. At zero temperature and integer filling factor, the
competition between kinetic energy and on-site repulsion induces
the MI-SF transition. When the on-site repulsion is dominating,
${t/U \ll 1}$ the atoms are tightly localized in the MI ground
state which is well approximated by the product of local (on-site)
Fock states. The Mott state is characterized by zero
compressibility originating from an energy gap for particle and
hole excitations. When the hopping amplitude is increased up to a
certain critical value ${(t/U)_{\rm c}}$,  particle delocalization
becomes energetically more favorable and the system Bose
condenses. In the chemical potential vs. hopping matrix element
plane (energies are scaled by $U$), the $T=0$ phase diagram has a
characteristic lobe shape ~\cite{Fisher}, see also
Fig.~\ref{phase_diagram} below, with the MI phase being inside the
lobe (there is one lobe for each integer filling factor). The most
interesting region in the phase diagram is the vicinity of the
lobe tip, $(\mu=\mu_{\rm c}, \, t=t_{\rm c})$, corresponding to
the MI-SF transition in the commensurate system. For other values
of $\mu$ or $t$, the SF-MI criticality is trivial and corresponds
to the weakly interacting Bose gas at vanishing particle density
\cite{Fisher}. It is straightforwardly described provided the
particle (hole) effective mass is known. If, however, one crosses
the MI-SF boundary at constant commensurate density (this is
equivalent to going through the tip of the lobe at a fixed
chemical potential) the long-wave action of the system becomes
relativistic and particle-hole symmetric. Now the phase transition
is in the four-dimensional U(1) universality class \cite{Fisher}.
It is worth emphasizing that here we have a unique opportunity of
a laboratory realization of  the non-trivial relativistic vacuum,
a sort of a ``hydrogen atom" of strongly-interacting relativistic
quantum fields. Approaching the critical point from the MI side,
one deals with the vacuum that supports massive bosonic particles
and anti-particles (particles and holes). On the other side of the
transition, the SF vacuum supports massless bosons (phonons) that
do not have an anti-particle analog. In principle, one can
systematically study universal multiparticle scattering amplitudes
of the relativistic quantum field theory in the ultra-cold
"supercollider"!

The present study is focused on the three-dimensional (3D) system.
To the best of our knowledge, previous systematic studies of the
3D case were limited to the mean-field (MF) \cite{Fisher} and
perturbative methods \cite{Monien}. In Ref.~\cite{Monien}, the
authors utilized the strong-coupling expansion to establish
boundaries of the phase diagram in the $(\mu /U,\; t/U)$ plane.
This approach, based on the small ratio ${zt/U\ll 1}$, where $z=6$
is the coordination number for the simple cubic lattice, works
well only in the MI phase in the region far from the tip of the
lobe, where the insulating gap is larger then hopping, $\Delta/zt
> 1$. Close to the critical region, where $\Delta/zt \sim 1$, the
strong-coupling expansion is no longer valid. We present the
results of large-scale Monte Carlo (MC) simulations of the model
(\ref{BH}) by worm algorithm \cite{worm}. With precise data for
the single-particle Green function, we are able to carefully trace
the critical and close-to-critical  behavior of the system, and,
in particular, produce an accurate phase diagram in the region of
small insulating gaps $\Delta \ll t$. Though the corresponding
parameter range is quite narrow, it is crucial to clearly resolve
it to reveal the emerging particle-hole symmetry and relativistic
long-wave physics at the tip of the MI-SF transition. We also
present data for the effective masses of particle and hole
excitations inside the insulating phase. Close to the MI lobe tip,
the data for the dispersion of the elementary excitations are
fitted by the relativistic law, in agreement with the theory (this
also allows us to extract the value of the sound velocity in the
critical region). In the Mott state, the knowledge of gaps and
effective masses is sufficient to calculate the partition function
in the low temperature limit analytically and to make reliable
predictions for the system entropy.

For such applications of the system as quantum information
processing and interferometry, controlling the temperature is of
crucial importance. Most applications are based on the key
property of the good insulating state, which is small density
fluctuations in the ground state. At zero temperature fluctuations
are of quantum nature and can be efficiently controlled externally
through the $t/U$ ratio. At finite temperature, fluctuations are
enhanced by thermally activated particle-hole excitations. Only
when the temperature is much smaller than the energy gap, the
number of excitations is exponentially small. Up to date, there
are no available experimental techniques to measure the
temperature of a strongly interacting system. For weakly
interacting systems, the temperature can be extracted in a number
of ways, e.g. from the interference pattern of matter waves
\cite{interference} or the condensate fraction observed after the
trap is released and the gas expands freely \cite{time_of_flight}
---these properties are directly related to the momentum
distribution function $n(\mathbf{k})$. For strongly interacting
systems, both temperature and interaction are responsible for
filling the higher momentum states, which makes it hard to extract
temperature using absorption imaging techniques.

The results presented in this paper can be used to perform
accurate thermometry. Typically, the initial temperature, $T^{\rm (in)}$,
(before the optical lattice is adiabatically loaded)  is
known. By entropy matching one can easily deduce the final
temperature of the MI state, $T=T^{\rm (fin)}$, provided the
entropy of the MI phase is known. To this end we have calculated
the energy, specific heat and entropy of the system in several
important regimes which include MI and strongly correlated SF
phases. These data can be used to suggest appropriate initial
conditions which make the Bose-Hubbard system suitable for
physical applications, such as the ones described above.

Another interesting question concerns the nature of inhomogeneous
states in confined systems when the MI phase is formed in the trap
center. The confining potential provides a scan in the chemical
potential of the phase diagram at fixed $t/U$ \cite{Jaksch}. As
one moves away from the trap center the system changes its local
state. At zero temperature, the density profile of the system can
be read (up to finite-size effects) from the ground state phase
diagram. At finite temperature, this is no longer possible. In
particular, the liquid regions outside of the MI lobes could be
normal or superfluid, depending on temperature.

\begin{figure}[t]
\centerline{\includegraphics[angle=0,scale=0.55] {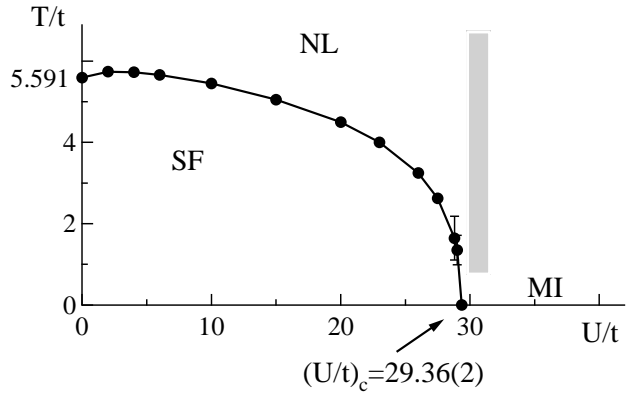}}
\caption{(Color online). Finite-temperature phase diagram at
filling factor $n=1$. Solid circles are simulation results (the
line is a guidance for the eye), error bars are plotted.
$T=5.591t$ is the critical temperature of the ideal Bose gas with
the tight binding dispersion relation. At finite, but low enough
temperature, the MI domain  is loosely defined as the part of the
phase diagram to the right of the gray line. The rest of the
non-superfluid domain is referred to as normal liquid (NL).}
\label{critical_T}
\end{figure}

So far experimental results have been interpreted by assuming that
liquid regions are superfluid, but there were no direct
measurements or calculations to prove that this was the case.
[Part of the problem is that absorption imaging is sensitive only
to $n(\bf k)$, which is the Fourier transform of the
single-particle density matrix in the relative coordinate. All
parts of the system contribute to $n(\bf k)$ and it is hard to
discriminate where the dominant contribution comes from.] It is
almost certain that $T^{\rm (fin)}/T_{\rm c}^{\rm (fin)}$ of the
strongly correlated system is higher then $T^{\rm (in)}/T_{\rm
c}^{\rm (in)}$. Indeed, since the entropy of MI at $\Delta \gg T$
is exponentially small, most entropy will be concentrated in the
liquid regions. At this point we notice that the transition
temperature in the liquid is suppressed relative to the
non-interacting Bose gas value $T_{\rm c}^{(0)}\, \approx\,
3.313\, n^{2/3}/m$ by both (i) effective mass enhancement in the
optical lattice, $m \to 1/2ta^2$ (here $a$ is the lattice
constant), and (ii) strong repulsive interactions in the vicinity
of the Mott phase, in fact, $T_{\rm c} \to 0$ at the SF-MI
boundary. It seems plausible that the MI phase is always
surrounded by a broad normal liquid (NL) region. It may also
happen that superfluidity is completely eliminated in the entire
sample in the final state. [Strictly speaking, at $T\ne 0$ the MI
and NL phases are identical in terms of their symmetries and are
distinguished only {\it quantitatively} in the density of
particle-hole excitations, i.e. in the Hamiltonian (\ref{BH}) the
finite-temperature MI is continuously connected without phase
transition to NL, see Fig.~\ref{critical_T}. For definiteness, we
will call NL a normal finite-$T$ state which is superfluid at
$T=0$ for the same set of the Hamiltonian parameters.]
Fig.~\ref{critical_T} shows the finite-temperature phase diagram
for filling factor $n=1$ (we will discuss how we determine the
critical temperature in Sec.~III). The critical temperature goes
to zero sharply, while approaching the critical point. In the
limit of $U\rightarrow 0$ the critical temperature is slightly
above the ideal-gas prediction ($T=5.591t$ was calculated using
the tight binding dispersion relation), as expected (see, e.g.
Ref.~\cite{critical_temp}).

The paper is organized as follows: in Sec. II we present results
for the ground state phase diagram and effective mass of particle
(hole) excitations, at integer filling factor $n=1$. In Sec. III
we investigate the thermodynamic properties of the system. We
present data for energy, specific heat and entropy and calculate
the final temperature of the uniform and harmonically confined
system in the limit of large gaps. For the case of trapped system,
we also determine the state of the liquid at the perimeter of the
trap. Brief conclusions are presented in Sec. IV.


\section{ground state properties}

This section deals with the results of large-scale Monte Carlo
simulations for the ground state phase diagram of the Bose-Hubbard
system in three dimensions. Analytical approaches, e.g. the strong
coupling expansion, work well in the region where $zt/U \ll 1$ and
the system is deep in the MI phase. Under these conditions the
kinetic energy term in the Hamiltonian can be treated
perturbatively and the unperturbed ground state is a product of
local Fock states. In Ref.~\cite{Monien} the authors carried out
an expansion, up to the third order in $zt/U$, for the SF-MI
boundaries and estimated positions of critical points at the tips
of the MI lobes (by extrapolating results to the infinite
expansion order). Their results agree with the mean field solution
calculated in Ref.~\cite{Fisher}, when the latter is expanded up
to the third order in $zt/U$ and the dimension of space goes to
infinity. As already mentioned, this approach starts failing when
$\Delta \sim zt$. Using MC techniques  we were able to calculate
critical parameters and predict the position of the diagram tip
with much higher accuracy: with the worm algorithm (WA) approach
the energy gaps can be measured with precision of the order of
$10^{-2} t$ \cite{chains99}. The simulation itself is based on the
configuration space of the Matsubara Green function
\begin{equation}
G(i,\tau) \; =\;   \langle \: {\cal T}_{\tau}  \:
b^{\dag}_{i}(\tau )\,  b_0^{\:}(0) \: \rangle \; , \label{G}
\end{equation}
which is thus directly available. We utilize the Green function to
determine dispersion relations for particle and hole excitations
at small momenta [from the exponential decay of $G(\textbf{p},
\tau )$ with the imaginary time]  which directly give us the
energy gap  and effective masses.

\begin{figure}[t]
\centerline{\includegraphics[angle=0,width=3.3in,height=2.9in]{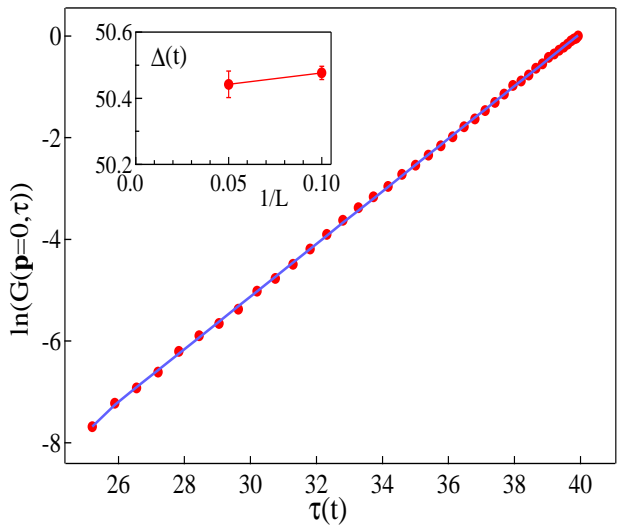}}
\caption{(Color online). Zero-momentum Green function in the Mott
phase with the chemical potential $\mu/U =0.809$, slightly below
the upper phase boundary. Here we show data for the system with
$N=10^3$ bosons at $U/t=70$ and $T/t=0.025$. In the inset we plot
the energy gap $\Delta$ for linear system sizes L=10 and L=20.
Finite-size errors are within the statistical error bars.}
\label{green_function}
\end{figure}

Recall that in the momentum space the Green function of a finite
size system $G(\textbf{p},\tau)$ is different from zero only for
$\textbf{p}=\textbf{p}_{\bf m} = 2\pi(m_x/L_x,\: m_y/L_y,
\:m_z/L_z )$, where $L_{\alpha=x,y,z}$ is the linear system size
in direction $\alpha$ (we performed all simulations in the cubic
system with $L_{\alpha}=L$), and $~\mathbf{m}=(m_x,m_y,m_z)~$ is
an integer vector. Using Lehman expansion and extrapolation to the
$\tau  \to \pm \infty$ limit one readily finds that

\begin{equation}
G(\mathbf{p},\tau)\; \to\; \left\{
\begin{array}{l}
Z_+e^{-\epsilon_+(\mathbf{p})\tau
}\; , ~~~~~~~\tau \, \to\, +\infty \; , \\
Z_-e^{\epsilon_-(\mathbf{p})\tau }\; , ~~~~~~~~~\tau \, \to\,
-\infty  \; .\end{array} \right. \label{G2}
\end{equation}
The two limits describe single-particle/hole excitations in the MI
phase. Here $Z_{\pm}$ and $\epsilon_{\pm}$ are the
particle/hole spectral weight (or $Z$-factors) and energy,
respectively. In the grand canonical ensemble, excitation energies
are measured relative to the chemical potential. With this in
mind, calculating the phase diagram of the system is rather
straightforward. At a fixed number of particles $N=L^3$ and $t/U$
ratio one determines chemical potentials $\mu_{\pm}$ for which the
energy gap for creating the particle/hole excitation with
$\mathbf{p}=0$ vanishes. The insulating gap is given then by
$\Delta = \mu_+ -\mu_-$. For high precision simulations of the gap
one has to choose the value of $\mu$ very close to $\mu_{\pm}$ and
consider finite, but zero for all practical purposes, value of
temperature so that the following two conditions are satisfied:
\begin{equation}
|\mu-\mu_{\pm}| \ll t\;, \;\;\;\;\;\; |\mu-\mu_{\pm}| \gg T  \;.
\label{conditions}
\end{equation}
This is exactly how we proceed. By plotting $\ln
[G(\mathbf{p},\tau)]$ vs. $\tau$ we deduce
$\epsilon_{\pm}(\mathbf{p})$ from the exponential decay of the
Green function. A typical example is shown in
Fig.~\ref{green_function}. We use the values of the hopping
amplitude $t$ and the lattice constant $a$ as units of energy and
distance, respectively.

\begin{figure}[t]
\centerline{\includegraphics[angle=0,width=3.3in,
keepaspectratio=true]{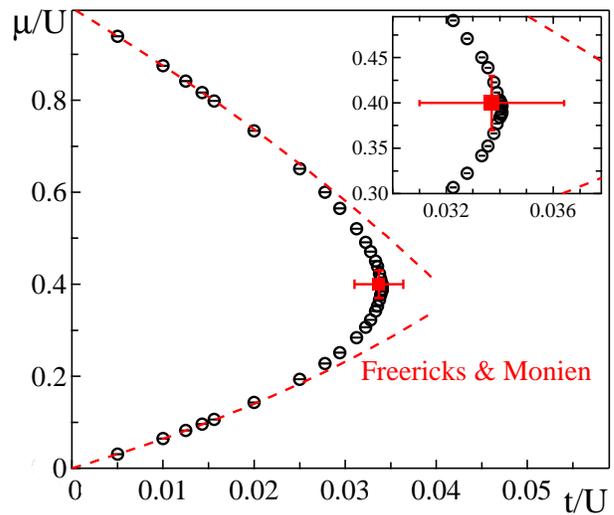}}
\caption{(Color online). Phase diagram of the first MI-SF lobe.
Numerical data are shown by open circles. The error bars
are shown but are barely visible even in the inset.
The dashed lines and the square represent results of
Ref.~\cite{Monien} for the strong coupling expansion and the
extrapolated position of the diagram tip, respectively.}
\label{phase_diagram}
\end{figure}


In Fig.~\ref{phase_diagram} we present accurate results for the
boundaries of the first Mott insulator lobe. We have done
calculations for systems with linear sizes $L= 5,\: 10,\: 15,\: 20$. Up
to values of $t/U\sim 0.031$ no size effects were detected within
the error bars. [Here and throughout the paper error bars are
of two standard deviations]. In the critical region the
finite-size effects were eliminated using standard scaling
techniques (see below). The dashed lines are the prediction of
Ref.~\cite{Monien} based on the third-order expansion in $t/U$. It
becomes inaccurate quite far from the tip when the insulating gap
is about $\sim 6t$. On the other hand, the value of the tip
position extrapolated to the infinite order is right on target,
within the error bar of order $3t$ for the chemical potential and
on-site repulsion. In all simulations (performed at $t/T=40$) the
finite-temperature effects are negligible---the system is
essentially in its ground state.

To eliminate finite-size effects in the critical region and
pinpoint the position of the lobe tip, we employed standard
scaling techniques based on the universality considerations.

First, let us briefly review the universal properties of the
insulator-to-superfluid transition (see Ref.~\cite{Fisher} for
more details). There exist two types of transitions: the
``generic" transition, when the phase boundary is crossed at fixed
\emph{t/U}, and a special transition at fixed integer density,
when the SF-MI boundary is crossed at fixed ${\mu/U}$. The generic
transition is driven by the addition/subtraction of a small number
of particles, and is fully characterized by the physics of the
weakly-interacting Bose gas formed by the small incommensurate
density component $n-n_0$, where $n_0$ is the nearest integer to
$n$. In particular, if $\delta $ is the deviation from the generic
critical point in the chemical potential or $t/U$ ratio then
$|n-n_0|\sim \delta$ and $T_{\rm c}(\delta ) \sim \delta ^{2/3}$
in the SF phase.

The special transition at the tip of the lobe happens at fixed
integer density. It is driven by delocalizing quantum fluctuations
which for large values of {\em t}/U enable bosons to overcome the
on-site repulsion and hop within the lattice. As explained in
Ref.~\cite{Fisher}, the effective action for the special
transition belongs to the $(d+1)$-dimensional XY universality
class which implies emergent relativistic invariance (rotational
invariance in the imaginary-time--space, which is equivalent to
the Lorentz invariance in real-time--space), and, in particular,
an emergent particle-hole symmetry. The upper critical dimension
for this transition is $(d+1)=4$, so that for $d>3$ the critical
exponents for the order parameter, $\beta $, and the correlation
length, $\nu$, are of mean-field character: $\beta=\nu=1/2$ (with
 logarithmic corrections for $d=3$). In this study, we were not able to resolve
logarithmic renormalizations for realistic 3D systems and proceed
below with the analysis which assumes mean-field scaling laws.

\begin{figure}
\hspace*{-0.4cm}
\includegraphics[width=3.0in,height=2.3in]{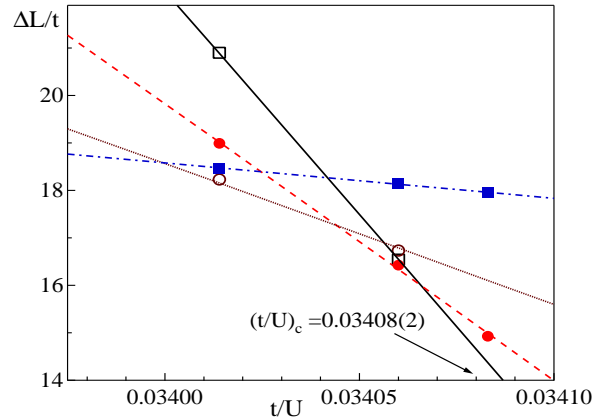}
\caption{(Color online). Finite size scaling of the energy gap at
the tip of the lobe.  $\Delta L/t$ vs. $t/U$ for system size $L$=5
(solid squares), $L$=10 (open circles), $L$=15 (solid circles),
$L$=20 (open squares). Lines represent linear fits used to extract
the critical point.} \label{critical_point}
\end{figure}

\begin{figure}
\hspace*{-1cm}
\includegraphics[width=3.0in,height=2.3in]{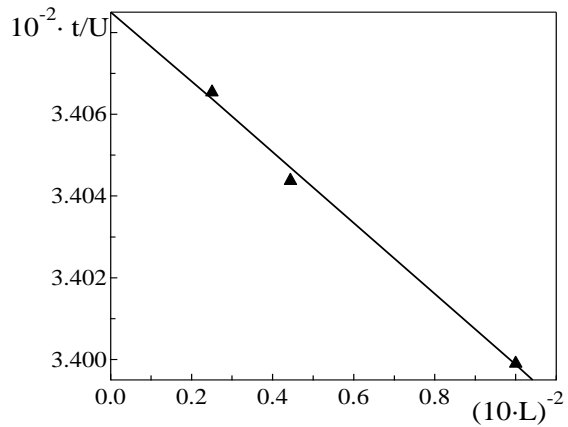}
\caption{ Extrapolation to the thermodynamic limit. We show the
intersections (triangles) of the curves (L=5, L=10), (L=10, L=15),
(L=15, L=20), vs. $L_{\rm max}^{-2}$. The fit (solid line) yields
$(t/U)_{\rm c}=0.03408(2)$.} \label{critical_point_extrap}
\end{figure}

Denoting the distance from the critical point as
$\gamma=[(t/U)_{\rm c}-t/U]$, for a system of linear size L one
can write
\begin{equation}
\Delta (\gamma ,L)\; =\;  \xi^{-1} f(\xi/L) \; =\;  L^{-1}g(\gamma
L^2) \; , \label{delta}
\end{equation}
where $\Delta$ is the particle-hole excitation gap, $\xi$ is the
correlation length, and $f(x)$ and $g(x)$ are the universal
scaling functions. In the last expression
we have used the relation  $\xi \propto \gamma^{-1/2}$. At the
critical point, the product $L\Delta $ does not depend on the
system size. Therefore, by plotting $L\Delta $ as a function of
$t/U$ one determines the critical point from the intersection of
curves referring to different values of L, as shown in
Fig.~\ref{critical_point}. This analysis yields
(Fig.~\ref{critical_point_extrap} explains how finite-size effect
in the position of the crossing point  originating from corrections
to scaling was eliminated)
\begin{equation}
(t/U)_{c}\; =\; 0.03408(2)~~~~~~~~~~~~(n\; =\; 1) \; . \label{}
\end{equation}
Our final results are summarized in
Fig.~\ref{phase_diagram}. We find that the size of the critical
region where 4D XY scaling laws apply is narrow and restricted to
small gaps of the order of $\Delta \le t$ (inside the vertical
error bar on the strong-coupling expansion result in
Fig.~\ref{phase_diagram}). It appears that resolving this limit
experimentally would be very demanding.

To perform analytic estimates of the MI state energy (and entropy)
at low temperature $T \ll \Delta $ one has to know effective
masses of particle and hole excitations, $m_{\pm}$. For example,
the particle/hole contributions to energy in the grand canonical
ensemble are given by the sums
\begin{equation}
E_\pm\;  =\;  \sum_{\mathbf k} \epsilon_\pm ({\mathbf k})\:
n_{\epsilon} \approx \left( {L \over 2\pi } \right)^{3} \int
d\mathbf{k}\: \epsilon_\pm ({\mathbf k}) \:e^{-\epsilon_\pm
({\mathbf k})/T}\; , \label{E1}
\end{equation}
where $n_{\epsilon}$ is the Bose function and
\begin{equation}
\epsilon_\pm({\mathbf k})\; \approx \; \pm(\mu_\pm-\mu)
+k^2/2m_\pm \label{dispersion}
\end{equation}
For large gaps the tight binding approximation
\begin{equation}
\epsilon_\pm({\mathbf k})\; \approx\;  \pm \, (\mu_\pm-\mu)\, +\,
\sum_{\alpha=x,y,z}\,{1-\cos k_\alpha\over m_\pm}
\label{dispersion2}
\end{equation}
is a reasonable approximation for all values of ${\bf k}$ in the
Brillouin zone. Note that, if one is to use the local density
approximation (LDA) for the energy/entropy estimates of trapped
systems, then calculations have to be performed in the grand
canonical ensemble.

\begin{figure}
\vspace*{-0.1cm}
\centerline{\includegraphics[angle=0,width=3.4in,keepaspectratio=true]{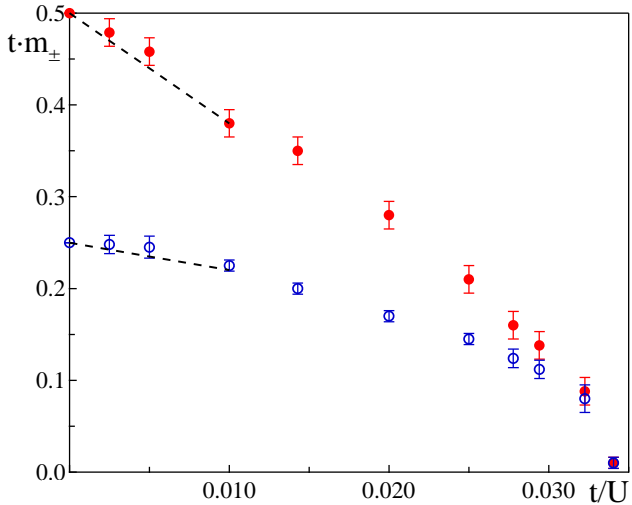}}
\caption{(Color online). Effective mass for hole (solid circles)
and particle (open circles) excitations as a function of $t/U$.
The exact results at $t/U=0$ are $m_{+} =0.25/t$ and $m_{-}
=0.5/t$. By dashed lines we show the lowest order in $t/U$
correction to the effective masses. Close to the critical point
the two curves overlap, directly demonstrating the emergence of
the particle-hole symmetry. At $t/U=0.034$, the sound velocity is
$c/t=6.3\pm 0.4$.}

\label{eff_mass}
\end{figure}

To determine effective masses we computed $G(\mathbf p , \tau)$ in
the insulating state and deduced $\epsilon_\pm ({\mathbf p})$ for
several lowest momenta from the exponential decay of the Green
function on large time scales. Dispersion laws were then fitted by
a parabola, with the exception for the diagram tip, where the
dispersion relation is relativistic. The result for $m_\pm$ is
shown in Fig.~\ref{eff_mass}. When $t/U \rightarrow 0$ one can
calculate effective masses perturbatively in $t/U$ to get
\begin{equation}
t\, m_{+} \; =\;  0.25 - 3t/U \; ,~~~~t\,  m_{-}\; =\;  0.5 -
12t/U \; . \label{masses}
\end{equation}
Clearly, our data are converging to the analytical result as
$\emph{t/U}\rightarrow 0$. On approach to the critical point the
effective mass curves become identical for particles and holes
indicating that there is an emergent  particle-hole symmetry at the diagram tip.
In agreement with the theoretical prediction, the data taken at
$\emph{t/U}=0.034$ are fitted best with the relativistic
dispersion relation $\varepsilon(p)=c\sqrt{m_*^2 \, c^2 +p^{2}}$,
where $c$ is the sound velocity and the effective mass is defined
as $m_*=\Delta/2c^2$. At this value of $t/U$ we have found that
$c/t\; =\; 6.3\pm 0.4$ and $ t\, m_*\; =\; 0.010\pm 0.004$.

\section{Finite temperature analysis}

Controlling the temperature is an important experimental issue,
crucial for many applications of cold atomic systems and studies
of quantum phase transitions. In this section, we discuss
thermodynamic properties of the Bose-Hubbard model. We present
data for energy, specific heat and entropy, for some specific
cases. In particular, we focus on the most important $\langle
n\rangle \, \approx\,  1$ situation. Our data can be used in two
ways: (i) to understand limits of applicability of the
semi-analytic approach (with calculated effective parameters)
discussed above, and (ii) to have reference first-principle curves
for more refined numerical analysis. Unfortunately, a direct
simulation of a realistic case in the trap, i.e. with similar
number of particles as in experiments, is still a challenging
problem though simulations of about $10^5$ particles or more at
low temperature seem feasible in near future.

The results are organized as follows: in Subsection A we probe the
limits of applicability of semi-analytic predictions in the Mott
state. In Subsection B we calculate the entropy of the
Bose-Hubbard model, compare it to the initial entropy (i.e. before
the optical lattice is turned on) and estimate the final
temperature.  We consider both  a homogeneous system in the MI
and SF states and a system of $N\sim 3\cdot10^4$ particles in a trap.

\subsection{Comparison with low T semi-analytic predictions}
Away from the tip of the lobe, in the MI state, semi-analytic
predictions are reliable provided the temperature is low enough,
i.e. $T\ll\Delta$. In other words, there exist a range of
temperatures , defined as $T\lesssim const\cdot \Delta$, where the
quasi-particle excitations can be successfully described as a
non-interacting classical gas (see Eq. (\ref{E1})). The value of
the constant depends on $\Delta$, as the two following examples
demonstrate.

\begin{figure}
\includegraphics[width=1.1\columnwidth,keepaspectratio=true]{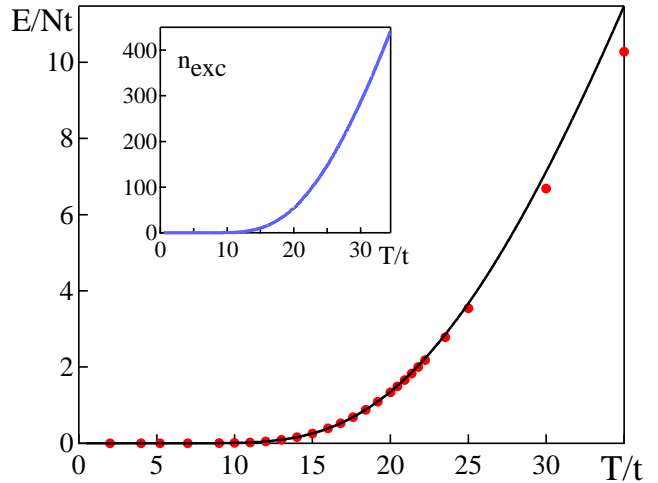}
\caption{(Color online). Energy per particle at $t/U=0.005$, unity
filling factor and linear system size $L=20$. Solid circles are
prime data (error bars within symbol size), the solid line is the
analytical prediction from Eq. \ref{E1}, where, at each
temperature, the chemical potential has been fixed by imposing
equal number of particle and hole excitations. The inset shows the
total number of excitations present in the system.}
\label{energy_U100_uniform}
\end{figure}
\begin{figure*}[htb]
\hspace*{-0.3cm}
\includegraphics[width=0.34\textwidth,keepaspectratio=true]{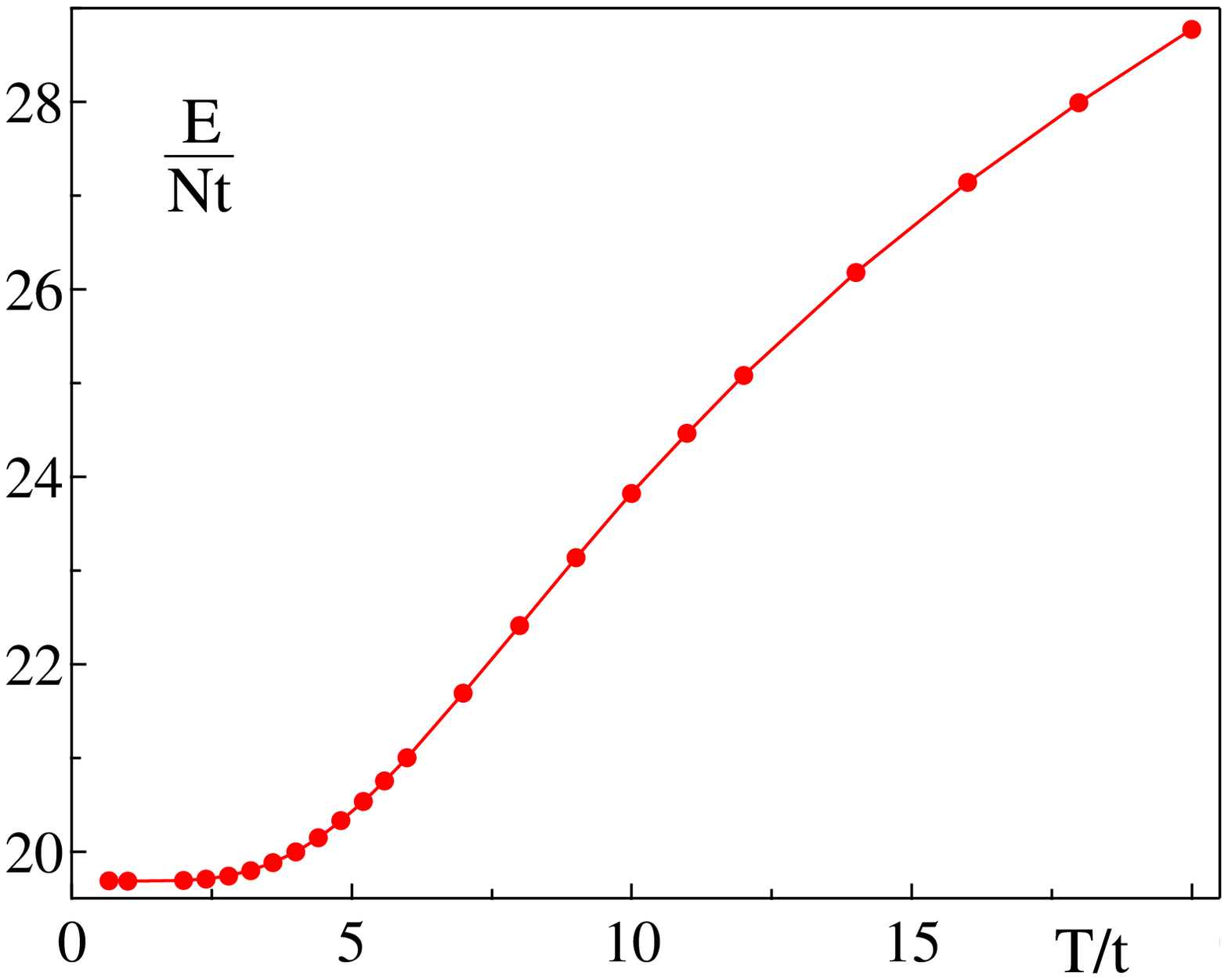}
\hspace*{-0.28cm}
\includegraphics[width=0.34\textwidth,keepaspectratio=true]{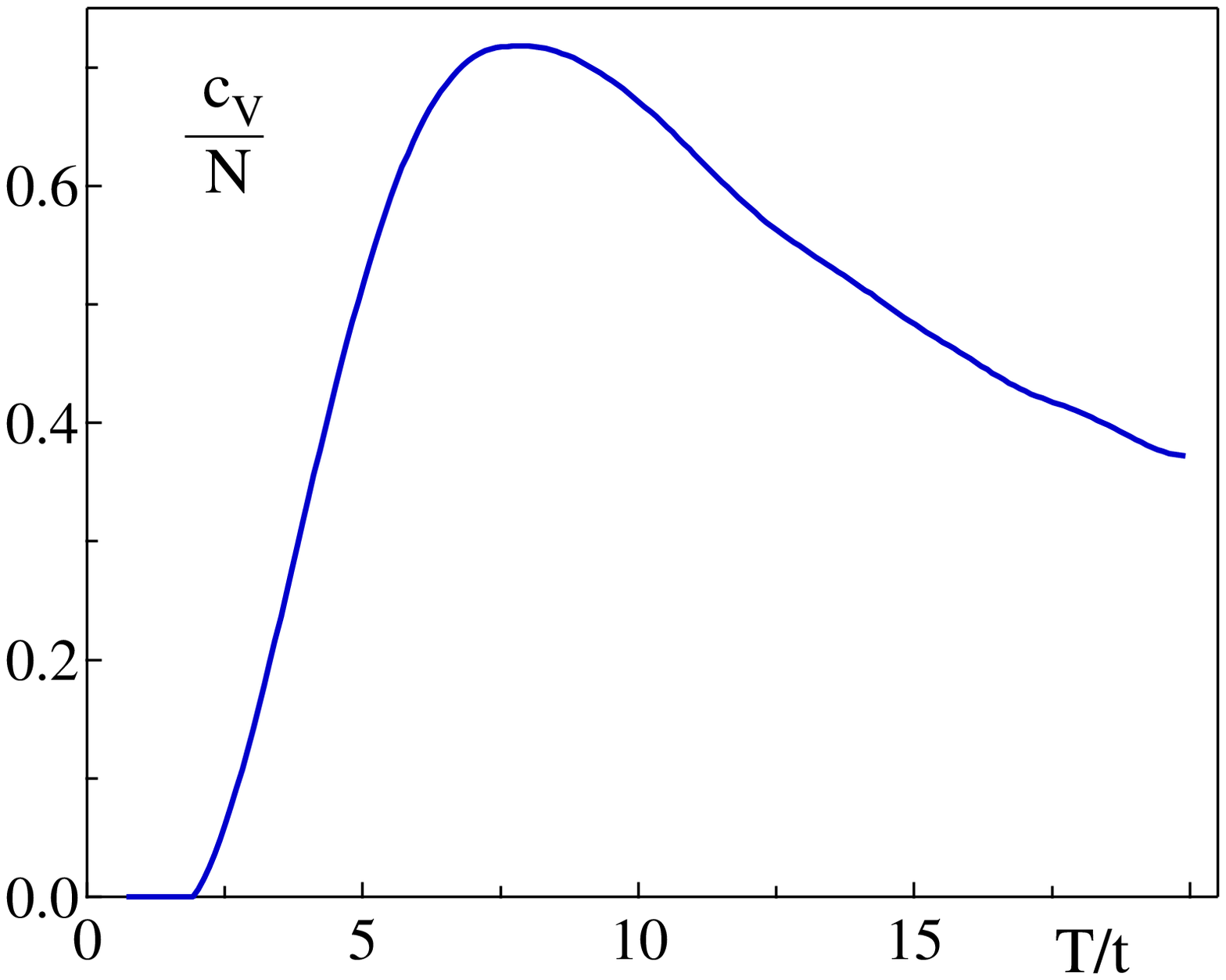}
\hspace*{-0.28cm}
\includegraphics[width=0.34\textwidth,keepaspectratio=true]{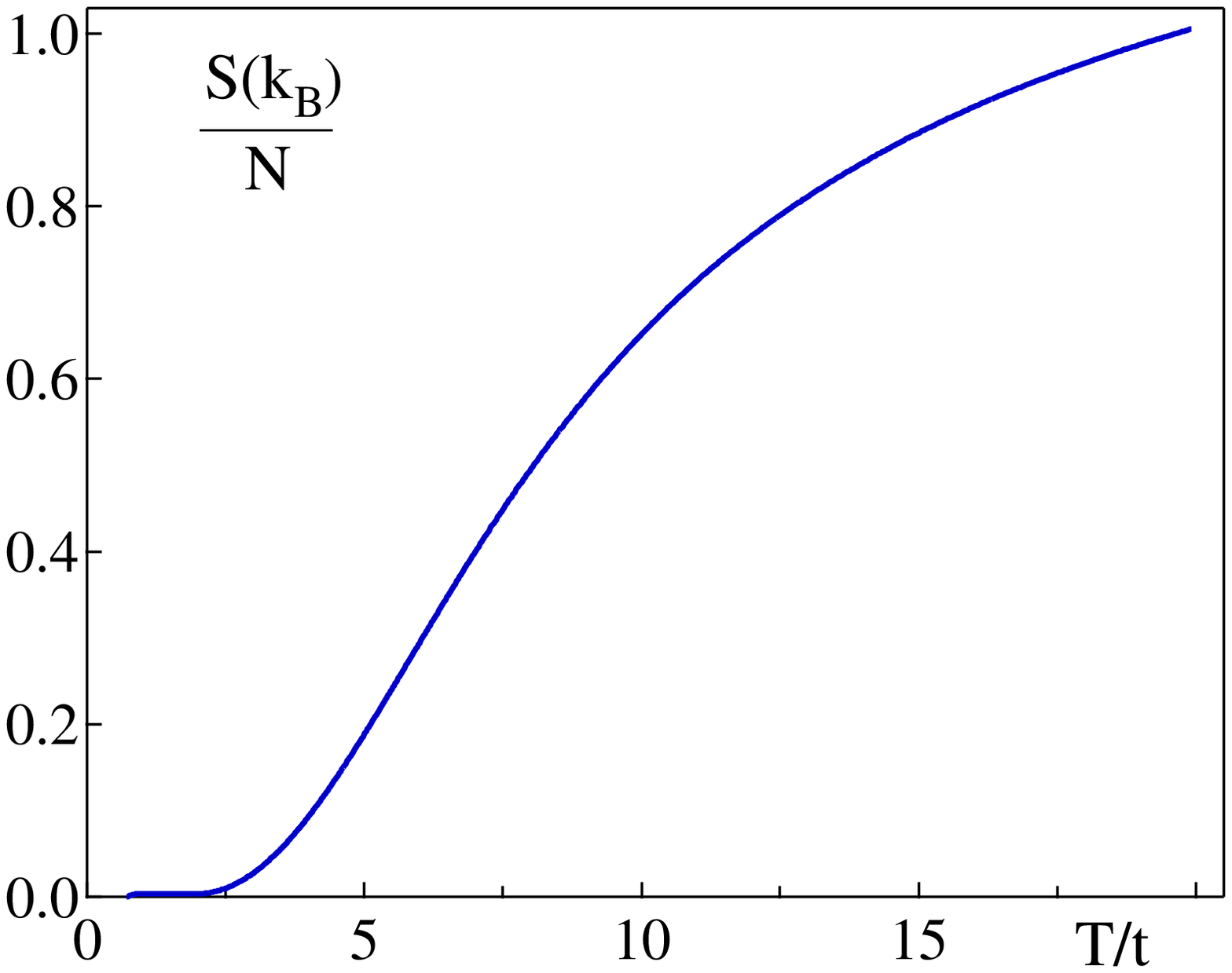}
\caption{(Color online). Energy (left), specific heat (center) and
entropy (right) per particle at $t/U=0.025$ (MI ground state) and
unity filling factor. On the left, solid circles refer to prime
data (error bars within symbol size). the data were taken for linear
system sizes L=10 and L=20. Within the error bars, we are not able to
resolve any finite size effect. Solid lines in all plots are
obtained from spline-interpolated data for energy, with subsequent
analytic differentiation/integration of the interpolation curve.}
\label{entropy40}
\end{figure*}

Let us first consider larger gaps, e.g. $\Delta \sim 200 t$, for
which we have found that the low temperature analytic predictions
reproduce numerical data very well. In Fig.
\ref{energy_U100_uniform} we plot the energy per particle in the
low temperature regime for $\Delta = 181.6 t$. The analytic
prediction from Eq. (\ref{E1}), where, at any given temperature,
the chemical potential has been chosen by setting equal the total
number of particle and hole excitations (as it is done for
intrinsic semiconductors), is reliable up to temperatures
$T\lesssim 35t$, i.e. $T\lesssim 0.15\Delta$. In the inset we plot
the average number of particle-hole excitations. This number
increases rapidly with temperature justifying the grand-canonical
calculation for the quasi-particle gas (at fixed total number of
particles). The quasi-particle number density is $\sim 5\%$ at
$T\sim35t$. Apparently, for higher temperatures the ideal gas
picture is no longer valid as it crosses over to that of the
strongly correlated normal liquid. We conclude that for large
enough gaps and $T\lesssim 0.15\Delta$, one can rely on low
temperature analytical predictions to do thermometry.

For smaller gaps, instead, we do not find any interesting region
(i.e. where temperature effects are visible), for which the
classical description is valid. In Fig.~\ref{entropy40} we show
the energy per particle as a function of temperature for
$t/U=0.025$ (the groundstate is MI with the energy gap $\Delta =
18.35 t$). To get the specific heat and entropy, we first use
spline interpolation of the energy data points to obtain a smooth
curve $E(T)$. The specific heat is then obtained by
differentiating the spline. The maximum in the specific heat is
reached when temperature is about half the energy gap. The entropy
has been calculated by numerical integration of $c_{V}/T$. In
order to see any temperature effect one has to go as high as
$T\sim 2.5t$; at these temperatures the classical description is
already no longer applicable and one has to rely on numerical data
to do thermometry.
\subsection{Loading the optical lattice: estimate of $T^{(\rm fin)}$ from entropy
matching}

The standard approach to convert results obtained for a
homogeneous system into predictions for systems in external fields
is the so-called local density approximation (LDA), which is
actually a local chemical potential approximation when the density
at the site $i$ is identified with the density of the homogeneous
system with the chemical potential
\begin{equation}
\mu_{i}^{\rm (eff)}\; =\; \mu \, - \, V(i)\; . \label{effective
mu}
\end{equation}
In strongly interacting regimes with a short healing and
correlation length, the LDA approach can be easily justified in
most cases (critical regions of phase transitions excluded). In
Ref.~\cite{Wessel}, the authors directly compare simulation
results for 1D and 2D harmonically trapped systems with LDA
predictions based on known homogeneous system phase diagram. As
expected, the density profiles differ only at the MI-SF interface,
and we believe that the same will be true for the 3D case which is
more ``mean-field-like".

When the  semi-analytic predictions are reliable (see Fig.
\ref{energy_U100_uniform}), one can use numerical results for the
effective masses and gaps to calculate the entropy of the
homogeneous quasi-particle gas with the
tight-binding dispersion relation. The entropy is given by:
\begin{equation}
S\; =\; -\frac{V}{(2\pi)^{3}}\int d^{3}\mathbf{k}\,
\frac{\partial[\Omega_{+}(\mathbf k)+\Omega_{-}(\mathbf
k)]}{\partial T}\; , \label{entropy_Mott}
\end{equation}
where
\begin{equation}
\Omega_{\pm}(\mathbf k)\; =\; T\, \ln \left(
1-\exp\frac{\epsilon_{\pm}(\mathbf k)}{T}\right) \; .
\label{Gibbs_energy}
\end{equation}

\begin{figure}
\vspace*{-0.3cm} \hspace*{-0.5cm}
\includegraphics[width=1.15\columnwidth,keepaspectratio=true]{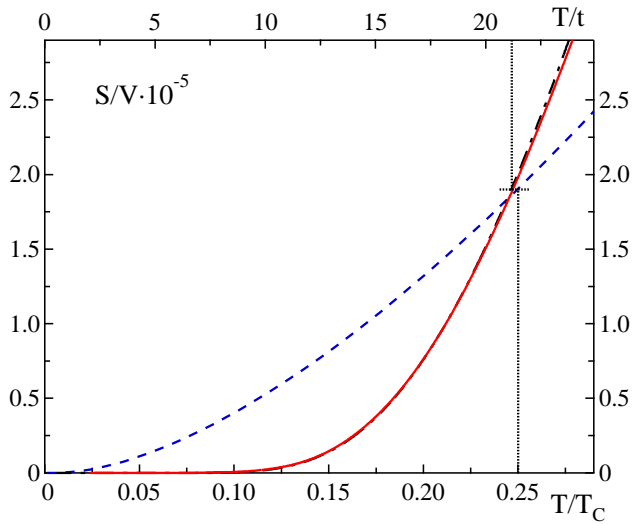}
\caption{(Color online). Entropy per particle at $t/U=0.005$,
unity filling factor and linear system size $L=20$. The dashed
line is the entropy of the uniform WIBG. The solid line is the
result of analytical derivation/integration of numerical data for
energy and the dashed-dotted line (barely visible) is the
analytical prediction,  Eq.~(\ref{entropy_Mott}),
where, at each temperature, the chemical potential has been fixed
by the condition of having equal number of particle and hole excitations. If the
system was initially cooled down to $T^{\rm in}=0.25T_{\rm c}$,
the final temperature is $T^{\rm fin}=22t$ and nearly a hundred of
thermally activated particle-hole excitations are present in the
final state (see inset in Fig.~\ref{energy_U100_uniform}).}
\label{entropy_uniform}
\end{figure}

As an example,  consider  a uniform weakly interacting Bose gas
(WIBG) of $^{87}$Rb with the gas parameter $na_{s}^{3}\sim
10^{-6}$, which is loaded into an optical lattice with
$\lambda=840$nm and $t/U=0.005$. At low enough temperature, $T\lesssim
0.3T_{\rm c}$, one can calculate the initial (prior to imposing
the lattice) entropy of the system using the Bogoliubov spectrum.
In Fig.~\ref{entropy_uniform} we plot the entropy per unit volume
before and after the optical lattice is imposed. The bottom
\emph{x} axis is temperature in units of the critical temperature
of the WIBG, the top \emph{x} axis is temperature in units of
$\emph{t}$, the hopping matrix element. The dashed line is the
entropy of the WIBG, the solid and dashed-dotted lines represent
the entropy of the Bose-Hubbard model calculated starting from the
numerical results of Fig.~\ref{energy_U100_uniform} and
analytical predictions of Eq.~(\ref{entropy_Mott}), respectively.
If the system was prepared at $T\sim 0.25T_{\rm c}$, the final
temperature would be $T\sim22t$. Fig.~\ref{energy_U100_uniform}
shows that at this temperature the system is quite far from its
ground state and the number density of thermally activated
particle-hole excitations is $\sim1\%$. The circumstances of this
kind become crucial if one is to use the system in quantum
information processing. This example is also illustrative of how
numerical data can be used to suggest appropriate initial
conditions.
%

Now we turn to a more realistic case of confined system and use
LDA to convert results for the uniform system into
predictions for the inhomogeneous one. In what follows, we
consider a gas of $N\sim3\cdot10^{4}$ $^{87}$Rb atoms,
magnetically trapped in isotropic harmonic potential of frequency
$2\pi60$ Hz. Experiments with such number of particles were
recently performed \cite{Gerbier1}. With this geometry, the
parameter $\eta =1.57(N^{1/6}a_{s}/a_{ho})^{2/5}$ (see
Ref.~\cite{Giorgini}) is $\sim 0.33$, which is a typical value in
current experiments. For temperatures in the range $\mu<T<T_{\rm
c}$, where $T_{\rm c}$ is the critical temperature of the
harmonically trapped ideal gas, one can accurately calculate
energy using the Hartree-Fock \cite{Hartee-Fock} mean field approach
\cite{Stringari}:
\begin{equation}
\frac{E}{NT_{c}}=\frac{3\zeta(4)}{\zeta(3)}t^{4}+\frac{1}{7}\eta(1-t)^{2/5}(5+16t^{3}),
\label{energy_trap}
\end{equation}
where \emph{t} is the reduced temperature $T/T_{\rm c}$. At very
low T ($T<\mu$), Eq.~(\ref{energy_trap}) misses the contribution
coming from collective excitations. We are interested in initial
temperatures $T\sim0.2-0.3 T_{\rm c}$, which are feasible  in
current experiments \cite{Gerbier}. Starting from
Eq.~(\ref{energy_trap}), we  calculate the entropy of the BEC
initially prepared in the magnetic trap. After the optical lattice
is adiabatically turned on, the magnetic potential provides a scan
over the chemical potential of the homogeneous system [see
Eq.~(\ref{effective mu})].

\begin{figure}
\includegraphics[width=1.0\columnwidth,keepaspectratio=true]{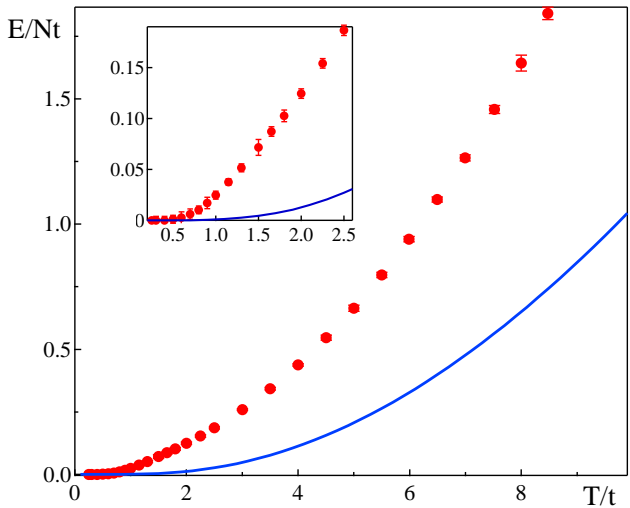}
\caption{(Color online). Energy per particle at $t/U=0.005$,
$\mu=116.5t$ and trap frequency $\omega=2\pi 60 Hz$. Solid circles
are numerical data, the solid line is the energy of particle and hole
 excitations in MI deduced from Eq. (\ref{E1}). The
inset shows a zoom of the low temperature range. At $T\sim2.3t$
the contribution of MI excitations to energy is $\sim 10\%$ only.}
\label{energy_harmonic}
\end{figure}

A direct comparison with experiments at fixed number of particles
would require to calculate $\mu(T)$ from the normalization
condition. At low temperatures, one expects the dependence of the
chemical potential on temperature to be weak (this will be
confirmed by direct simulations, see below). For simplicity, we
fix $\mu$ at a value corresponding to $N=30^{3}$ trapped atoms in
the first Mott lobe, at zero temperature. From this point we
proceed in two directions. On one hand, we analytically calculate
the low temperature contribution to  energy and entropy arising
from particle and hole excitations in the trapped MI state.
On the other hand, we directly simulate the thermodynamics of
the inhomogeneous system at a fixed chemical potential.
The results are shown in Fig.~{\ref{energy_harmonic}}, where we plot
the energy per particle, counted from the ground state. The solid
circles are data from the simulations (error bars are plotted),
the solid line is the (analytically calculated) contribution of
the particle and hole excitations. The inset shows the low
temperature region. A large mismatch between the two results
indicates that the main contribution to energy is given by the
liquid at the perimeter of the trap. At zero temperature, there
are about $N\sim29000$ particles in the trap, $7\%$ of which are not in the
MI state (recall that $\mu$ has been determined by placing
$30^{3}$ particles in the MI state). Simulation results show
that, in the range of temperatures considered, the total number of
particles increases by $0.7\%$ at most, which confirms the weak
temperature dependence of the chemical potential. In addition, for
$T=8t$, we performed a simulation with $N$ fixed at the {\it
groundstate} value. The energies per particle in the canonical and
grand canonical simulations differ by $0.3\%$ only, therefore we
proceed calculating the entropy in the canonical ensemble and
compare it with the initial entropy of the system, at a fixed
particle number.

\begin{figure}
\vspace*{-0.2cm}
\includegraphics[width=1.0\columnwidth,keepaspectratio=true]{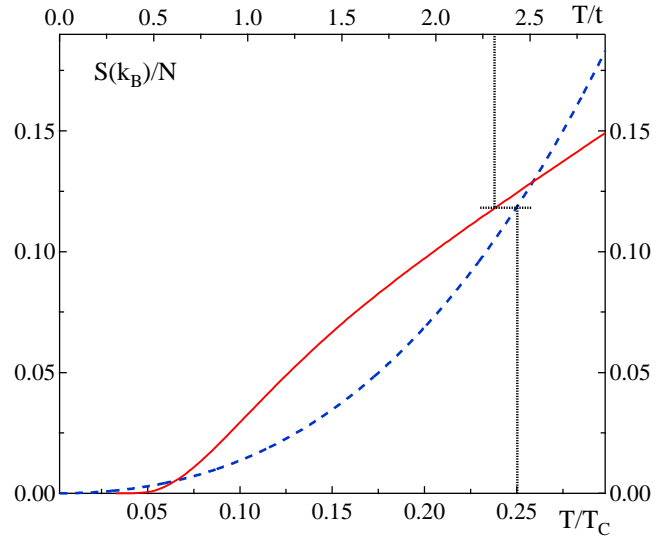}
\caption{(Color online). Entropy per particle for the same system
as in Fig.\ref{energy_harmonic}. The dashed line is the entropy of
the trapped WIBG, before the optical lattice is loaded; the solid
line is the final entropy. If the system was initially cooled down
to $T^{(\rm in)}=0.25T_{\rm c}$, the final temperature is $T^{(\rm
fin)}=(2.35\pm0.3)t$ and the liquid at the perimeter of the trap
is normal (see text). } \label{entropy_trap}
\end{figure}

We are in a position to address the question of what is the final
temperature of the system after the optical lattice is turned on
and the final state is MI with the exception of a small shell at
the trap perimeter. In Fig.~\ref{entropy_trap} we plot the entropy
of the trapped WIBG with (solid line) and without (dashed line)
the optical potential. If the initial system is cooled down to
temperatures $T^{(\rm in)}\sim 0.25T_{\rm c}$, see, e.g.,
Ref.~\cite{Gerbier}, the final temperature will be $T^{(\rm
fin)}=(2.35\pm0.30)t$.

With the initial conditions considered in this example, what is
the final state of the liquid at the perimeter of the trap? Before
answering this question, we would like to recall that, along the
MI-SF transition lines, the critical temperature for the
normal-to-superfluid transition is zero. The transition
temperature increases as one moves away from the border of the
Mott lobes (lowering $\mu$ at fixed $t/U$ in our case) and the
quasi-particle density increases until it reaches its maximum at
about $n\approx 1/2$ and then decreases. The maximum $T_{\rm c}$
can be estimated from the ideal Bose gas relation $T^{(0)}_{\rm
c}=3.313n^{2/3}/m^{*}=4.174t$, with $n=0.5$ and $m^{*}=1/2t$, but
interaction effects are likely to reduce this value. We have
performed simulations at half filling factor and fixed
$t/U=0.005$, and found the critical temperature to be $T_{\rm
c}(n=1/2)=2.09(1)t$. As a consequence, for the chosen initial
conditions, we can conclude that the final state of the liquid at
the perimeter of the trap is normal and it gives the main
contribution to the entropy. For such low final temperature, the
contribution to the entropy per particle due to thermally
activated excitations in the MI state is only 10\%. The largest
chemical potential is, in fact, deep in the first Mott lobe, and
the energy required to introduce an extra particle or hole is much
larger than $T$. Most excitations are located at the perimeter of
the Mott state in a narrow shell of radius $R$ and width $\sim
0.05R$.
%

\begin{figure*}[htb]
\hspace*{-0.45cm}
\includegraphics[width=0.34\textwidth,keepaspectratio=true]{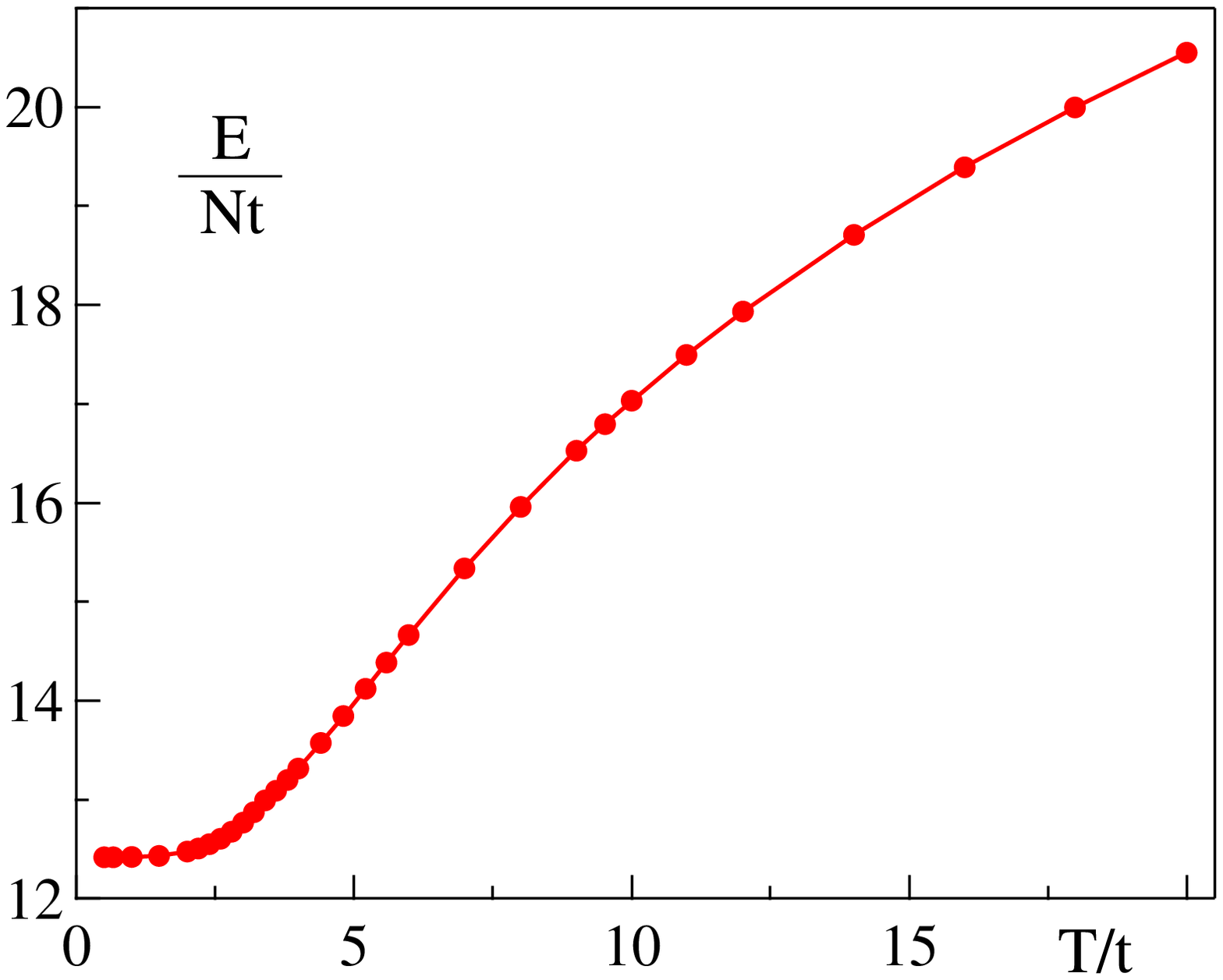}
\hspace*{-0.32cm}
\includegraphics[width=0.34\textwidth,keepaspectratio=true]{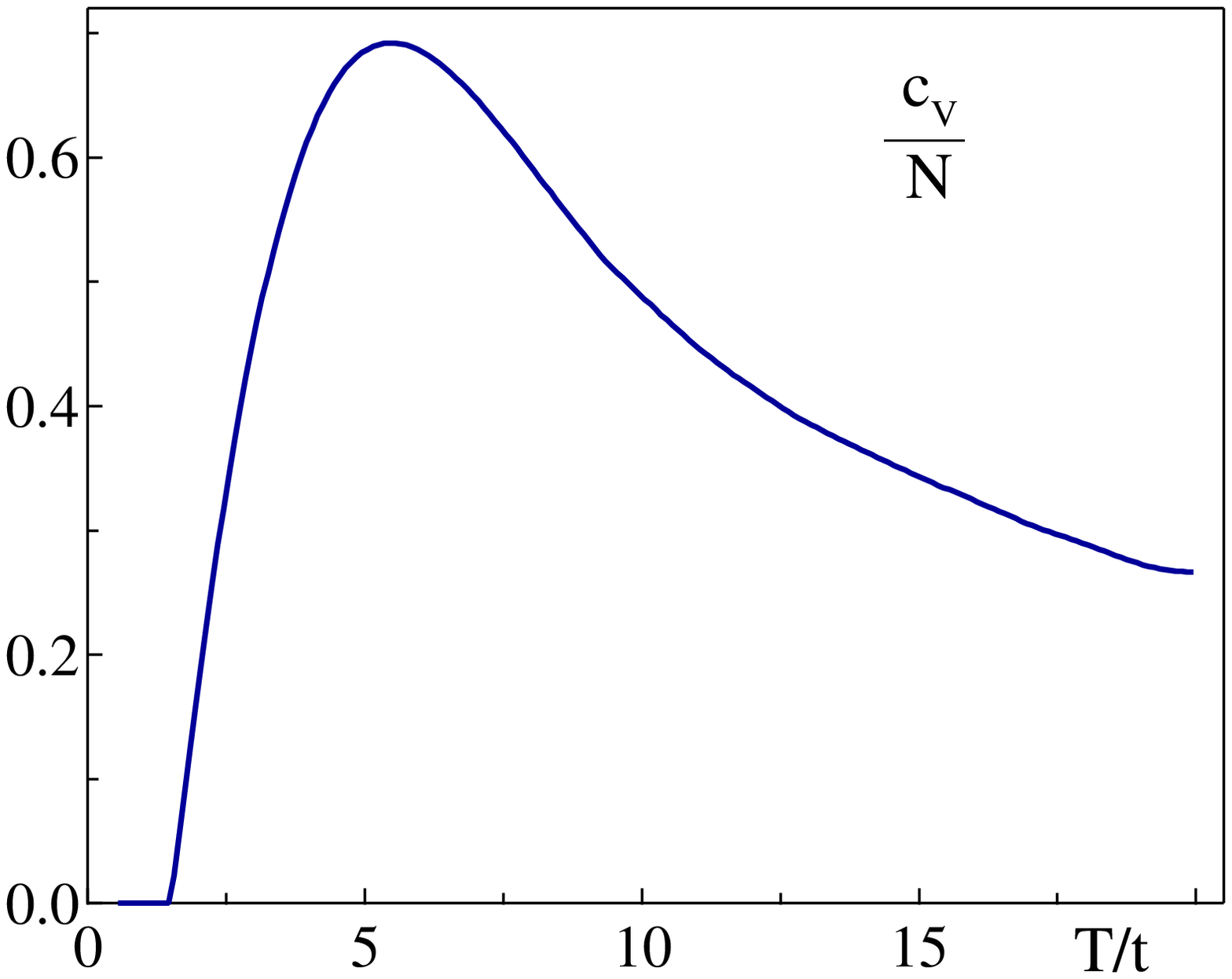}
\hspace*{-0.33cm}
\includegraphics[width=0.34\textwidth,keepaspectratio=true]{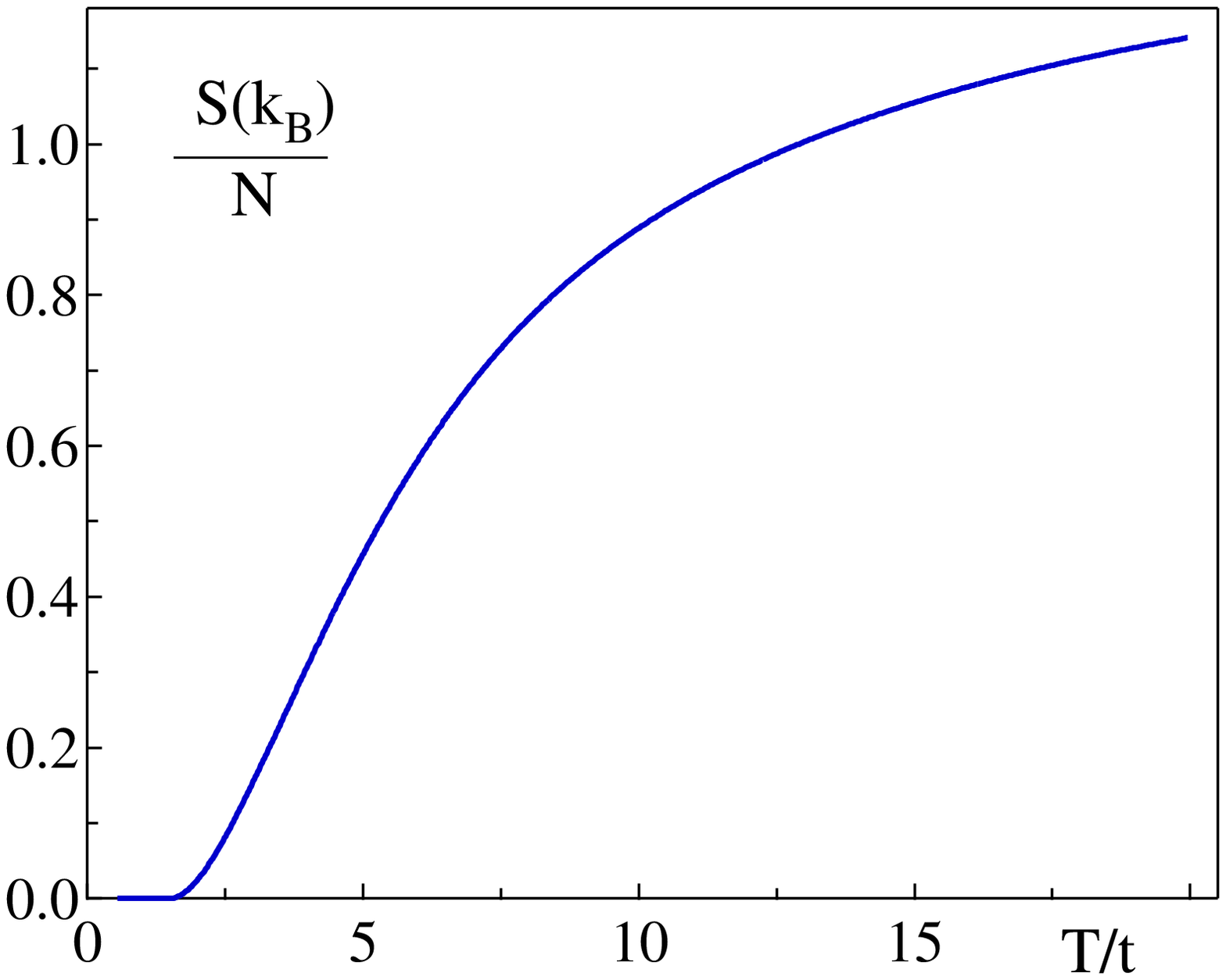}
\caption{(Color online). Energy (left), specific heat (center) and
entropy (right) per particle at $t/U=0.0385$ (SF ground state) and
unity filling factor. On the left, solid circles refer to prime
data (error bars within symbol size). Data were taken for system
size L=10 and L=20. Within error bars, we are not able to resolve
any finite size effect. Solid lines in all plots are obtained from
spline-interpolated data for energy, with subsequent analytic
differentiation/integration of the interpolation curve.}
\label{entropy26}
\end{figure*}

Retrieving the same information for experiments using a larger
number of particles, e.g. $10^{5}\div 10^{6}$, by direct
simulation is still computationally challenging. In order to use
LDA, one should study the uniform system, scanning through the
chemical potential. As our last example, we consider a uniform
system which is in the correlated SF ground state.
Fig.~\ref{entropy26} shows data for $E$, $c_{\rm V}$, and $S$ for
$t/U=0.0385$ and unity filling factor, close to the MI-SF
transition. The system stays in its ground state for $T \ll 2t$;
in finite systems, the energy of the lowest mode is finite:
$E_{\rm min}=cp_{\rm min}$, with $c\thickapprox 6t$ and $p_{\rm
min}=2\pi/L$). The specific heat and entropy are calculated as
described for Fig.~\ref{entropy40}. We were not able to resolve
the SF-NL transition temperature from this set of data alone:
numerical data corresponding to system sizes $L=10$ and $L=20$
overlap within error bars and we did not see any feature at the
critical temperature. This is not surprising since the specific
heat critical exponent $\alpha$ is very small, and it is thus very
difficult to resolve the singular contribution and finite size
effects in energy and specific heat. For a system of linear size
$L$, the singular part of the specific heat, $c_{\rm V}^{(\rm
s)}$, can be written as
\begin{equation}
c_{V}^{(s)} (t ,L)\;  =\;  \xi^{\alpha/\nu} f_c(\xi/L)\;  =\;
L^{\alpha/\nu}g_c(t L^{1/\nu}) \; , \label{spec_heat scaling}
\end{equation}
where $t=(T-T_{c})/T_c$, $\alpha\approx-0.01$, $\nu=(2-\alpha)/3$,
$\xi$ is the correlation length and $f_c(x)$ and $g_c(x)$ are the
universal scaling functions. At the critical point, finite size
effects for the two system sizes considered are $\sim 1\%$, within
error bars.
\begin{figure}
\includegraphics[width=0.99\columnwidth,keepaspectratio=true]{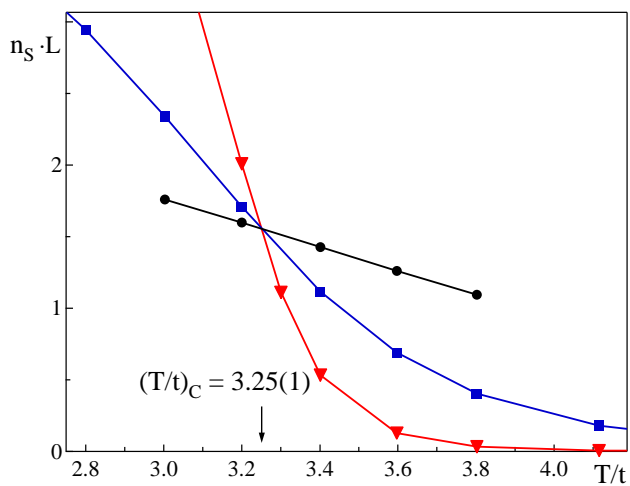}
\caption{(Color online). Finite size scaling of the superfluid
stiffness. $n_s L$ vs. $T/t$ for system size $L$=5 (circles),
$L$=10 (squares), $L$=20 (triangles). We estimate the critical
temperature to be $T_{\rm c}=3.25(1)t$, already nearly half the
non-interacting gas value.}
\label{SF_density}
\end{figure}

The critical temperature was extracted from data for the
superfluid stiffness. The scaling of the superfluid stiffness at
the critical temperature is $n_s\varpropto|t|^{\nu}$. This allows
one to accurately estimate the critical temperature from
\begin{equation}
n_{s} (t ,L) = \xi^{-1}f_s(\xi/L)\;  =\; L^{-1}g_s(t L^{1/\nu}) \;
, \label{SF_density scaling}
\end{equation}
by plotting $n_sL$ vs. $T$ as shown in Fig.~\ref{SF_density}. From
the data taken for system sizes $L=5$, $L=10$, $L=20$, we estimate
the critical temperature to be $T_c=3.25(1)t$. At this
temperature, the entropy per unit particle is $\sim0.195$, or,
translating to entropy density in physical units,
$3.6\cdot10^{-5}JK^{-1}m^{-3}$, which corresponds to an initial
temperature $\sim0.35T_{\rm c}^{(\rm in)}$. Therefore it seems
plausible to reach $T_{\rm c}$ experimentally.

\section{Conclusions}
We have performed quantum Monte Carlo simulations of the three
dimensional homogeneous Bose-Hubbard model. We were able to
establish the phase diagram of the MI-SF transition with the
record accuracy $\sim 0.1{\%}$ and determine the size of the
fluctuation region in the vicinity of the diagram tip where
universal properties of the relativistic effective theory can be
seen. Comparison with the strong-coupling expansion shows that the
latter works well only for sufficiently large insulating gaps
$\Delta > 6t$ outside of the fluctuation region. We have studied
the effective masses of particle and hole excitations along the
MI-SF boundary. Our results directly demonstrate the emergence of
the particle-hole symmetry at the diagram tip, and provide base
for accurate theoretical estimates of the MI thermodynamics at low
and intermediate temperatures.

We have studied thermodynamic properties of the superfluid and
insulating phases at fixed particles number for the uniform case.
These data can be used to make predictions for the inhomogeneous
system using the local density approximation. We have shown that
for large enough gaps the low temperature analytical predictions
agree with numerical data. By entropy matching, we have calculated
the final temperature of the system (after the optical lattice is
adiabatically loaded), in the uniform and magnetically trapped
system, at $t/U=0.005$. We have performed direct simulations of a
trapped system, using typical experimental values for the magnetic
potential and number of particles. For the initial conditions
considered, we found the final temperature and demonstrated that
the main contribution to the entropy comes from the liquid at the
perimeter of the trap. We have calculated the normal-to-superfluid
transition temperature at the half filing and concluded that the
liquid at the perimeter is normal.

\section{Acknowledgements}

We are grateful to Matthias Troyer for useful discussions.
This  work was supported by the National Science Foundation Grant
No. PHY-0426881.

 \end{document}